\documentclass[11pt]{article}
\newenvironment{numberedlist}
{\begin{list}{\makebox[20pt]{\hss(\arabic{itemno})\enspace}}
             {\usecounter{itemno}\labelwidth 20pt}}{\end{list}}

\newcounter{itemno}

\newcounter{itemno1}

\newcounter{itemno2}
\newcounter{lemma}
\newcounter{exno}

\newcounter{defno}

\newenvironment{defn}{\refstepcounter{defno}\medskip \noindent {\bf
Definition \thedefno.\ }}{\medskip}

\newcommand{\sep}{\;\vert\;}

\newcommand{\oprove}{\vdash\kern-.6em\lower.7ex\hbox{$\scriptstyle O$}\,}

\newcommand{\Pscr}{{\cal P}}

\newcommand{\pderivation}{{\cal P}\kern -.1em\hbox{\rm -derivation}}
\newcommand{\pderivationl}{{\cal P}\kern -.1em\hbox{\em -derivation}}
\newcommand{\pderivable}{{\cal P}\kern -.1em\hbox{\rm -derivable}}
\newcommand{\pderivablel}{{\cal P}\kern -.1em\hbox{\em -derivable}}
\newcommand{\pderivations}{{\cal P}\kern -.1em\hbox{\rm -derivations}}
\newcommand{\pderivability}{{\cal P}\kern -.1em\hbox{\rm -derivability}}

\newcommand{\all}{\forall}
\newcommand{\some}{\exists}

\newsavebox{\lpartfig}
\newsavebox{\rpartfig}

\newenvironment{exmple}{
 \begingroup \begin{tabbing} \hspace{2em}\= \hspace{3em}\= \hspace{3em}\=
\hspace{3em}\= \hspace{3em}\= \hspace{3em}\= \kill}{
 \end{tabbing}\endgroup}

\newcommand{\lb}{\langle}
\newcommand{\rb}{\rangle}

\newcommand{\Ra}{\supset}  
\newcommand{\adc}{\&} % choice conjunction

\newtheorem{theorem}[lemma]{Theorem}

     {\\* \hspace*{\fill} \end{trivlist}}

\newcommand{\prov}{pv}

\begin{document}

\begin{center}
{\Large {\bf  Mutually Exclusive Rules in LogicWeb
}}
\\[20pt] 
{\bf Keehang Kwon and Daeseong Kang}\\
Dept. of Computer Engineering and Electrical Engineering, DongA University \\
Busan 604-714, Korea\\
%051-200-7784 \\
\{ khkwon,dskang \}@dau.ac.kr\\
\end{center}

\noindent 
LogicWeb has traditionally lacked devices for expressing 
 mutually exclusive clauses. We address this limitation 
  by adopting 
   choice-conjunctive clauses of the form 
 $D_0 \adc D_1$  where  $D_0, D_1$ are Horn clauses and 
 $\adc$ is a linear logic connective.   Solving a  goal $G$ using  $D_0 \adc D_1$ -- $\prov(D_0 \adc D_1,G)$ -- has the 
 following operational semantics: choose a successful one between $\prov(D_0,G)$ and
 $\prov(D_1,G)$. In other words, if $D_o$ is chosen in the course of solving $G$, then $D_1$ will be
discarded and vice versa.  Hence, the class of  choice-conjunctive clauses precisely captures
the notion  of  mutually exclusive clauses.

\section{Introduction}\label{sec:intro}

Internet computing is an important modern programming paradigm.
 One successful attempt
towards this direction is LogicWeb\cite{LD96}. LogicWeb is a model of the World Wide Web, where Web pages 
are represented as logic programs, and hypertext links represents logical implications
 between these programs.
 LogicWeb is  an integral part of Semantic Web\cite{Davies}. Despite much 
attractiveness, LogicWeb (and its relatives such as agent programming) has 
traditionally lacked elegant devices for structuring mutually exclusive rules.
 Lacking such devices, structuring mutually exclusive rules in LogicWeb relies on awkward devices such as 
the cut or $if$-$then$-$else$ construct\cite{Por}.

\newcommand{\hweb}{LinWeb}

This paper proposes \hweb, an extension to LogicWeb with a novel feature called choice-conjunctive clauses.
 This logic extends Horn clauses  by the choice construct 
 of the form $D_0 \adc D_1$ where $D_0, D_1$ are Horn clauses and $\adc$ is a choice-conjunctive connective of
linear logic.
Inspired by \cite{Jap03}, this has the following intended semantics: $choose$ a successful one between $D_0$ and $D_1$ in the course of 
solving a goal. 
This expression thus supports the idea of mutual exclusion. 

An illustration of this aspect is provided by the following clauses $c1,c2$ which define the usual
$max$ relation:

\begin{exmple}
$c1:  max(X,Y,X)  :-\ X\geq Y. $ \\
$c2:       max(X,Y,Y)  :-\ X < Y.$
\end{exmple}
\noindent 
 These two clauses are mutually exclusive. Hence, only one of these two clauses
can  succeed. Therefore, a more economical definition which consists of one clause  $c3$ is possible:

\begin{exmple}
$c3:      (max(X,Y,X)  :-\ X\geq Y) \adc$\\
\hspace{2em}      $(max(X,Y,Y)  :-\ X < Y). $
\end{exmple}
\noindent
This  definition is more economical (and more deterministic) in the sense that it reduces the search space by
cutting out the other alternatives.
For example, consider a goal $max(9,3,Max)$.
Solving this goal  has the effect of choosing  the first conjunct of (a copy of) $c3$, producing
the result $Max = 9$.
Our machine, unlike Prolog and other linear logic languages such as Lolli \cite{HM94}, does not create a
backtracking point for the second conjunct.
The key difference between our language and other logic languages is that the selection action is present in our
 semantics, while it is not present at all in other languages.

The remainder of this paper is structured as follows. We describe \hweb\
 in
the next section. In Section \ref{sec:modules}, we
present some examples of \hweb.
Section~\ref{sec:conc} concludes the paper.

\section{The Language}\label{sec:logic}

The language is an extended  version of Horn clauses
 with choice-conjunctive clauses. It is described
by $G$- and $D$-formulas given by the syntax rules below:
\begin{exmple}
\>$G ::=$ \>   $A \sep  G \land  G \sep   D \Ra  G \sep \some x\ G $ \\   \\
\>$D ::=$ \>  $A  \sep G \supset D\ \sep \all x\ D \sep D \adc D$\\
\end{exmple}
\noindent
In the rules above,   
$A$  represents an atomic formula.
A $D$-formula  is called a  Horn
 clause with choice-conjunctive clauses. 
 
In the transition system to be considered, $G$-formulas will function as 
queries and a set of $D$-formulas will constitute  a program. 

 We will  present an operational 
semantics for this language. The rules of \hweb\ are formalized by means of what it means to
execute a goal task $G$ from a program $\Pscr$.
These rules in fact depend on the top-level 
constructor in the expression,  a property known as
uniform provability\cite{Mil89jlp,MNPS91}. Below the notation $D;\Pscr$ denotes
$\{ D \} \cup \Pscr$ but with the $D$ formula being distinguished
(marked for backchaining). Note that execution  alternates between 
two phases: the goal-reduction phase (one  without a distinguished clause)
and the backchaining phase (one with a distinguished clause).

\begin{defn}\label{def:semantics}
Let $G$ be a goal and let $\Pscr$ be a program.
Then the notion of   executing $\lb \Pscr,G\rb$ -- $\prov(\Pscr,G)$ -- 
 is defined as follows:
\begin{numberedlist}

\item  $\prov(A;\Pscr,A)$. \% This is a success.

\item    $\prov((G_1\supset D);\Pscr,A)$ if 
 $\prov(\Pscr, G_1)$ and  $\prov(D;\Pscr, A)$.

\item    $\prov(\all x D;\Pscr,A)$ if   $\prov([t/x]D;
\Pscr, A)$.

\item    $\prov(D_0 \adc D_1;\Pscr,A)$ if  choose a successful disjunct between  $\prov(D_0;\Pscr, A)$ and
 $\prov(D_1;\Pscr, A)$. 

\item    $\prov(\Pscr,A)$ if   $D \in \Pscr$ and $\prov(D;\Pscr, A)$. \%  change to backchaining phase.
\item  $\prov(\Pscr,G_1 \land G_2)$  if $\prov(\Pscr,G_1)$  and
  $\prov(\Pscr,G_2)$.

\item $\prov(\Pscr,\exists x G_1)$  if $\prov(\Pscr,[t/x]G_1)$.

\item $\prov(\Pscr, D\Ra G_1)$ if $\prov(\{ D \}\cup \Pscr,G_1)$
% This goal behaves as exclusive-OR. 

\end{numberedlist}
\end{defn}

\noindent  
In the rule (4), the symbol $D_0 \adc D_1$  allows for the mutually exclusive execution of clauses. This rule
 can be implemented as follows:
  first attempts to solve the goal using $D_0$.
 If it succeeds, then do nothing (and do not leave any choice point for $D_1$
). If it fails, then $D_1$ is attempted.

The following theorem connects our language to linear logic.
Its proof is easily obtained from the discussions in  \cite{HM94}.

\begin{theorem}
 Let ${\cal P}$ be a program and 
let $G$ be a goal.  Then, $\prov(\Pscr,G)$ terminates with a success
 if and only if $G$ follows from
$\Pscr$ in intuitionistic linear logic. 
\end{theorem}

\section{\hweb}\label{sec:modules}

In our context, a web page corresponds simply to a set of $D$-formulas
 with a URL. 
The module construct $mod$ allows a URL to be associated to a set of $D$-formulas.
An example of the use of this construct is provided by the 
following ``lists'' module which contains some basic list-handling rules.

\begin{exmple}
 $mod(www.dau.com/lists)$.\\
%$htext(www.krx.com/graphdoc.html)$. \%  web page \\
%$path(X,Y)$ ${\rm :-}$ \> \hspace{4em} $edge(X,Y).$\\               
%$path(X,Y)$ ${\rm :-}$ \> \hspace{4em} $edge(X,Z), path(Z,Y).$\\
\%  deterministic version of the member predicate \\
$memb(X,[X|L])\ \adc $\\
$memb(X,[Y|L])$ {\rm :-}  $(neq\ X\ Y)\ \land\ memb(X,L).$\\
\%  optimized version of the append predicate \\
$      append([],L,L)\ \adc $ \\
$      append([X|L_1],L_2,[X|L_3])$ {\rm :-} $append(L_1,L_2,L_3).$  \\ 

\% the union of two lists without duplicates \\
$uni([],L,L)\ \adc $\\
$uni([X|L],M,N)$ {\rm :-}  $memb(X,M)\land uni(L,M,N) \adc $\\
$uni([X|L],M,[X|N]) $ {\rm :-} $uni(L,M,N)$.
\end{exmple}
Our language  makes it possible to change $memb$ to be deterministic and more efficient: 
only one occurrence can be
found. Our approach can be beneficial to most Prolog deterministic definitions. For example,
the above definition of $append$ explicitly tells the machine not
to create a backtracking point. This is in constrast to the  usual one in Prolog in which
mutual exclusion must be inferred by the Prolog interprter.

 These  pages can be made available in specific contexts by explicitly
mentioning the URL via a hyperlink. For example, consider a goal 
$www.dau.com/lists \Ra  uni([a,b],[b,c],Z)$. This goal is translated to
$D_1 \Ra D_2 \Ra \ldots uni([a,b],[b,c],Z)$ where each $D_i$ is a $D$-formula in the $lists$.
Solving this goal  has the effect
 of adding each rule in $lists$
to the program before evaluating $uni([a,b],[b,c],Z)$, producing the result $Z = [a,b,c]$.

\section{Conclusion}\label{sec:conc}

In this paper, we have considered an extension to Prolog with  
mutually exclusive  clauses. This extension allows clauses of 
the form  $D_0 \adc  D_1$  where $D_0, D_1$ are Horn clauses.
These clauses are 
 particularly useful for replacing the cut in Prolog, making Prolog
more efficient and more readable.  We are investigating the connection between \hweb\ and Japaridze's
computability logic \cite{Jap03,Jap08}.
%\profile*{}{}% without picture of author's face

\section{Acknowledgements}

This work  was supported by Dong-A University Research Fund.

\bibliographystyle{plainr}% bib style

%\profile*{}{}% without picture of author's face

\end{document}